\documentclass{elsart}

\usepackage{epsfig}

\usepackage{amssymb}

\begin{document}

\begin{frontmatter}



\title{Multifractality of Cloud Base Height Profiles} 


\author[label1]{K.~Ivanova}
\author[label2]{N.~Gospodinova}
\author[label1]{H.N.~Shirer}
\author[label3]{T.P.~Ackerman}
\author[label2]{M.A.~Mikhalev}
\author[label4]{M.~Ausloos}

\address[label1]{Department of Meteorology, Pennsylvania State University, 
University Park, PA 16802, USA}
\address[label2]{Institute of Electronics, Bulgarian Academy of Sciences, 
72 Tzarigradsko chaussee, Sofia 1784, Bulgaria} 
\address[label3]{Pacific Northwest National Laboratory, Richland, WA 99352, USA}
\address[label4]{SUPRAS and GRASP, B5, University of Li$\grave e$ge, 
B-4000 Li$\grave e$ge, Belgium}

\begin{abstract}
Cloud base height profiles measured with laser ceilometer are studied using 
multifractal approach. The irregular structure of the 
signals is a benchmark  for nonlinear dynamical processes. A hierarchy of 
generalized dimensions determines the intermittency of the signal. 
The multi-affine properties are described by the $h(\gamma)$ function. 
\end{abstract}

\end{frontmatter}

% main text
\section{Introduction}

Cloud base height profiles (CBHP) are known to have highly fluctuating and  
irregular structure. The dynamics of CBHP evolution is determined
by a variety of processes in the atmosphere \cite{garratt}. This irregular 
structure  is a benchmark for nonlinear dynamical processes. We apply the multifractal
approach to study the scaling properties of the CBHP.
In this report we present an analysis of the multi-affine properties of a 
CBHP signal $y(t)$ (Fig.1) measured with a
ground-based laser ceilometer\footnote{http://www.arm.gov} having a temporal resolution of 
30 seconds, taken on September 23-25, 1997 at the Southern Great Plains (SGP)
(Oklahoma, USA) site 
of the Atmospheric Radiation Mesurement Program of the Department of Energy. 
\begin{figure}[ht] \begin{center} \leavevmode \epsfysize=8cm \epsffile{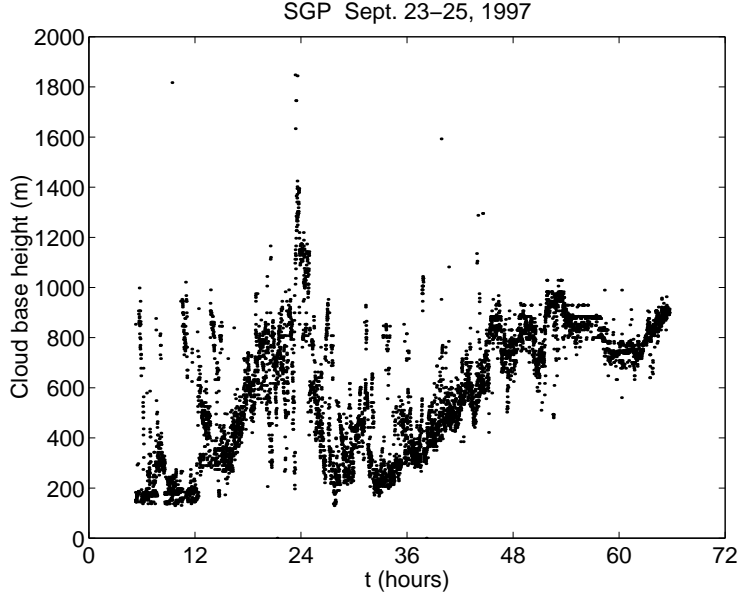}
\caption{Cloud base height profile data
measured at SGP site on Sept. 23-25, 1997. Time resolution :
30~sec; 7251 data points.}
\end{center} \end{figure}

\section{Multi-affinity and Intermittency} 

First, we tested the scaling properties of the power spectral density $S(f)$ 
of the CBHP signal and obtained (figure not shown) that $S(f)~f^{-\beta}$ 
with $\beta=1.46\pm0.06$ for frequencies lower than 1/15 min$^{-1}$ and 
$\beta\approx 0$ for higher frequencies.

The multi-affine properties
of $y(t)$ can be described by the so-called ``q-th'' order
structure functions \cite{davis,kita,kima2} 
$c_q = \left<|y(t_{i+r}) - y(t_i)|^q \right>$ ($ i=1,2, 
\dots, N -r $), 
where the averages are taken over all possible pairs of points that are  
$\tau=t_{i+r}-t_i$ apart from each other with $r>0$.  Assuming a power law dependence 
of the structure function, the $H(q)$ spectrum
is defined through \cite{davis} $ c_q(\tau)\sim \tau^{qH(q)}$ with 
$q\ge 0$.  

The {\it intermittency} of the signal can be studied through the so-called singular measure analysis. The first step that this technique requires is defining
a basic measure $\varepsilon(1;l)$ as
$\varepsilon(1;l)=|\Delta y(1;l)| / <\Delta y(1;l)>$ 
($l=0,1, \dots, N -1$), 
where $\Delta y(1;l)=y(t_{i+1})-y(t_i)$ is the small-scale gradient field and 
$<\Delta y(1;l)>=
\frac{1}{N}\sum_{l=0}^{N-1}|\Delta y(1;l)|. \label{ave}
$
It should be noted that we use spatial/temporal averages rather
than ensemble averages, thus making an ergodicity assumption \cite{ergodmf}
as our only recourse in such an empirical data analysis. Next we define a 
series of ever more coarse-grained and ever shorter fields $\varepsilon(r;l)$ 
where $0<l<N-r$ and $r=1,2,4,\dots,N=2^m$. The average measure in the 
interval $[l;l+r]$ is 
$\varepsilon(r;l)=\frac{1}{r}\sum_{l'=l}^{l+r-1} \varepsilon(1;l')$ 
($l=0, \dots , N - r$)
The
scaling properties of the generating function are then searched for through  
$\chi_q(\tau)=<\varepsilon(r;l)^q>\sim \tau^{-K(q)}$, for $q\ge0$. 
Thus the multi-fractal
properties
of the CBHP signal are expressed by two
sets of scaling
functions, $H(q)$ describing the roughness of the signal
and $K(q)$ describing its intermittency. The intermittency of the signal can 
be also expressed through the generalized dimensions $D(q)$ as introduced by 
Grassberger \cite{grass} and
Hentschel and Procaccia, \cite{proc}
$D(q) = 1 - K(q)/(q-1).
$
\begin{figure}[ht] \begin{center} 
\leavevmode \epsfysize=8cm \epsffile{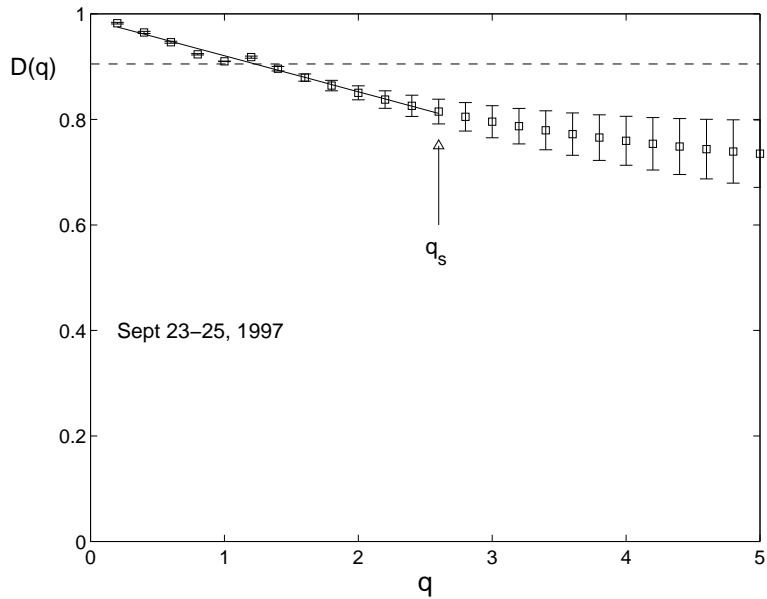}
\caption{Hierarchy of generalized dimensions $D(q)$ for the CBHP data in Fig.1. 
The straight line is drawn to enhance the value $q_s$ at which $D(q)$ function 
starts to deviate from a linear dependence. This deviation is related to 
sampling problems due to a single realization. The dashed line
defines the monofractal case $D_1=0.905$.} 
\end{center} \end{figure}

The multi-affinity of $y(t)$ means that one should use different scaling
factors $H(q)=H_q$ in order to rescale such a signal. This also implies that 
{\it local}
roughness exponents $\gamma$ exist \cite{nvmahg} at  different scales. 
The density of
the points $N_{\gamma}(\tau)$ that have the same roughness exponent is 
assumed \cite{mf}
to scale over the time span $\tau$ as $
N_{\gamma}(\tau)\sim \tau^-h(\gamma).$
 From Ref. \cite{bara} the following relations are found : 
$\gamma(q)=d(qH(q))/dq$
and 
$h(\gamma_q)=1+q\gamma(q)-qH(q).$
In Fig. 3 the CBHP multi-affine properties are presented via 
the $h(\gamma)$-curve.
\begin{figure}[ht] \begin{center} 
\leavevmode \epsfysize=8cm \epsffile{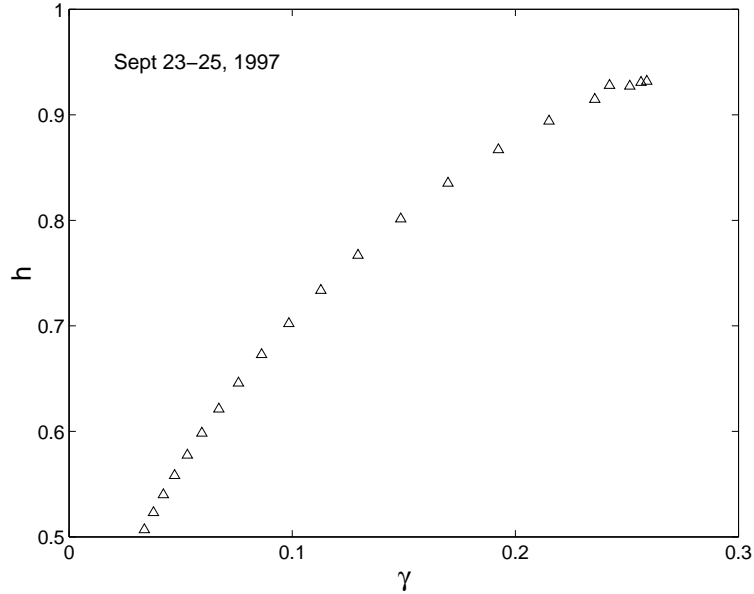}
\caption{The $h(\gamma)$-curve
for the CBHP data in Fig.1 }
\end{center} \end{figure}
\section{Conclusions}
We have demonstrated the multi-affine structure of cloud base height profiles.
Further work will be directed toward relating these statistical parameters to 
the dynamical properties of the clouds, an important step toward understanding,
modeling and predicting their dynamical behaviour.

\vskip 0.4cm

{\bf Acknowledgements} \vskip 0.2cm
This research was supported in part by the Department of Energy through grant
Battelle 327421-A-N4.

\end{document}